\documentclass[fleqn,usenatbib]{mnras}

\usepackage{textcomp,graphicx,amsmath,pdflscape,amssymb,setspace,subfig,rotating,multirow,psfrag,color}
\def\HeII{He\,{\scriptsize II}}

\def\OIII{[O\,{\scriptsize III}]}
\def\Hb{H$\beta$}

\title[Narrow-band \textit{HST} observations of SN 2014J]{Progenitor constraints on the Type Ia supernova SN 2014J from {\it Hubble Space Telescope} H$\beta$ and [O\,{\LARGE III}] observations}
\author[Graur \& Woods]
{Or~Graur$^{1,2}$\thanks{E-mail: \href{mailto:or.graur@cfa.harvard.edu}{or.graur@cfa.harvard.edu}}\thanks{NSF Astronomy and Astrophysics Postdoctoral Fellow.}
and Tyrone E. Woods$^3$\thanks{E-mail: \href{mailto:tewoods@star.sr.bham.ac.uk}{tewoods@star.sr.bham.ac.uk}}
\\
$^1$Harvard-Smithsonian Center for Astrophysics, 60 Garden St., Cambridge, MA 02138, USA \\
$^2$Department of Astrophysics, American Museum of Natural History, Central Park West and 79th Street, New York, NY 10024-5192, USA \\
$^3$Institute of Gravitational Wave Astronomy and School of Physics and Astronomy, University of Birmingham, Edgbaston, \\Birmingham B15 2TT, United Kingdom \\
}

\begin{document}

\maketitle


\setstretch{1}

\begin{abstract}
\noindent Type Ia supernovae are understood to arise from the thermonuclear explosion of a carbon-oxygen white dwarf, yet the evolutionary mechanisms leading to such events remain unknown. Many proposed channels, including the classical single-degenerate scenario, invoke a hot, luminous evolutionary phase for the progenitor, in which it is a prodigious source of photoionizing emission. Here, we examine the environment of SN 2014J for evidence of a photoionized nebula in pre- and post-explosion \OIII\ $\lambda5007$~\AA\ and H$\beta$ images taken with the {\it Hubble Space Telescope}. From the absence of any extended emission, we exclude a stable nuclear-burning white dwarf at the location of SN 2014J in the last $\sim$100,000 years, assuming a typical warm interstellar medium (ISM) particle density of 1~cm$^{-3}$. These limits greatly exceed existing X-ray constraints at temperatures typical of known supersoft sources. Significant extreme-UV/soft X-ray emission prior to explosion remains plausible for lower ISM densities (e.g., $n_{\rm ISM}\sim 0.1~\rm{cm}^{-3}$). In this case, however, any putative nebula would be even more extended, allowing deeper follow-up observations to resolve this ambiguity in the near future. 

\end{abstract}



\section{Introduction}
\label{sec:intro}

Type Ia supernovae (SNe Ia) have long been used to measure extragalactic distances and constrain cosmological models (and are responsible for creating a little more than half of the iron in our blood; \citealt{2017ApJ...848...25M}). Yet we still do not know what stellar evolutionary sequence leads to a SN Ia, although a number of channels have been proposed (see, e.g., \citealt{2018PhR...736....1L,2018SCPMA..61d9502S,2018RAA....18...49W} for theory reviews and \citealt{2014ARA&A..52..107M} for an observational review). In the following, we focus on testing the viability of progenitor scenarios that imply a hot, luminous phase prior to explosion, most notably including the classic single-degenerate (SD) scenario. 

In the SD scenario, a carbon-oxygen white dwarf (WD) accretes hydrogen from a giant or sub-giant companion \citep{Whelan1973}. As accretion continues, the density and temperature in the core of the WD rise until carbon is ignited in a thermonuclear runaway and the WD is disrupted. For the WD to efficiently grow in mass, the accreted hydrogen must undergo stable nuclear-burning on its surface. This means the progenitor system will be a luminous source of soft X-ray emission \citep[a supersoft X-ray source, SSS,][]{1992A&A...262...97V} for at least some period of time before the explosion. Note that steady nuclear-burning of helium (e.g., \citealt{2014MNRAS.440L.101R}) as well as some models for WD mergers (e.g., \citealt{2014A&A...563A..16N}) and possibly also post-merger surviving WDs \citep{2017ApJ...834..180S} predict a SSS phase as well. Dedicated searches for SSSs have failed to find a sufficient population of such systems to account for the observed SN Ia rates in the SD scenario in both young and old stellar populations \citep{2010ApJ...712..728D,2010Natur.463..924G,2016MNRAS.461.4505J}.

The extreme-UV photons from the SSS should also photoionize the warm ($\sim10^4$ K) interstellar matter (ISM) in the vicinity of the WD, creating a diffuse nebula that should be observable both \emph{before} and \emph{after} the SN Ia explosion \citep{1994ApJ...431..237R,2013MNRAS.432.1640W}. This is due to the low number densities ($n_{\rm{ISM}} \sim 1~\rm{cm}^{-3}$) and consequently long recombination times in the warm ISM, on the order of $10^5\times(1~{\rm cm^{-3}}/n_{\rm ISM})$~yr. Therefore, even if nuclear burning on the surface of the WD ceased tens of thousands of years prior to the explosion, nebulae should still be visible for thousands of years \emph{after} the SN \citep{2017NatAs...1..800W}. 

SN Ia explosions themselves are not significant sources of ionizing photons \citep{1978ApJ...225L..27C}. Indeed, spectra of SN 2011fe taken out to $\approx1000$ days after explosion show no sign of emission in \HeII, H$\beta$, or \OIII\ \citep{2015MNRAS.454.1948G}. The latter exclude progenitors up to temperatures of $T\sim10^4$~K, just below the temperature range where limits on photoionized nebulae become particularly constraining (a few $10^4$~K).

The study of emission-line nebulae ionized by nuclear-burning WDs is now on a firm theoretical footing \citep{1994ApJ...431..237R,2013MNRAS.432.1640W}, although detecting such objects requires deep flux sensitivity \citep{1995ApJ...439..646R,2014MNRAS.442L..28G,2016MNRAS.455.1770W}. \citet{2014MNRAS.442.1079J} searched the spectra of elliptical galaxies for \HeII\ $\lambda4686$~\AA\ (hereafter simply \HeII) line emission from ionization of the neutral gas by the integrated emission from a putative population of accreting, nuclear-burning WDs and found that the strength of this line was consistent with originating solely from the background population of post-asymptotic giant branch stars within the area covered by the fiber aperture of the Sloan Digital Sky survey, limiting the contribution of this channel to 5--10\% of the total SN Ia rate. Likewise, \citet{2014MNRAS.442L..28G} searched narrow-band \textit{Hubble Space Telescope} (\textit{HST}) F469N images of the nearby SN 2011fe taken before the explosion but found no trace of He{\scriptsize II} at the location of the SN.

The works cited above already present a stark problem for the standard SD scenario. However, only $\sim 0.1\%$ of the bolometric luminosity of the nebula is predicted to be emitted through the \HeII\ line. On the other hand, the \OIII\ $\lambda5007$~\AA\ (hereafter \OIII) line is predicted to be more than an order of magnitude stronger than the \HeII\ line. Though weaker than \OIII, H$\beta$ is still predicted to be roughly twice as strong as \HeII\ \citep{1994ApJ...431..237R}, while also directly tracing the recombination rate in the ISM, and allowing one to constrain lower source temperatures. Searching for extended \OIII\ and H$\beta$ emission in the vicinity of known SNe Ia can thus provide a powerful additional constraint on their progenitors \citep{2016MNRAS.455.1770W}.

SN 2014J, in the nearby galaxy M82, is the closest SN~Ia to have exploded in decades (3.3 Mpc; \citealt{2009ApJS..183...67D,2015ApJ...798...39M,2019ApJ...870...14G}). Other models for the SD scenario have already been severely constrained by SN 2014J. For example, \citet{2014ApJ...790....3K} used pre-explosion \textit{HST} images to exclude SD progenitor systems in which the companion is a bright red giant (including symbiotic novae comparable in luminosity to RS Oph). X-ray non-detections either limit the mass-loss rate from a SD companion to $\dot{M}<10^{-9}~{\rm M_\odot}~{\rm yr}^{1}$, assuming a wind velocity of 100 km~s$^{-1}$ or the density of the circumstellar material surrounding the SN to $<3~{\rm cm}^{-3}$ \citep{2014ApJ...790...52M}. Assuming the same wind velocity, radio non-detections constrain the mass-loss rate to $\dot{M}<7\times10^{-10}~{\rm M_\odot}~{\rm yr}^{1}$ and the circumstellar material density to $<1.3~{\rm cm}^{-3}$ \citep{2014ApJ...792...38P}. \citet{2014MNRAS.442.3400N} used {\it Chandra} pre-explosion images to rule out a SSS with the photospheric radius of a WD near the Chandrasekhar mass and a mass accretion rate in the narrow window where the hydrogen accreted onto the WD can burn stably. The {\it Chandra} constraint, however, only strictly applies to the time of the observations, on the order of years prior to explosion.

There are two major differences between the constraints set by these multi-wavelength observations and the ionization constraints presented here: (a) the radio and X-ray observations probe the interaction of the outflow from the progenitor with the ISM, while we probe the luminosity and effective temperature of the progenitor as it photoionizes the ISM; and (b) radio and X-ray observations can probe the state of the progenitor $10^3$--$10^4$ yr before the explosion, while our observations reach as far back as $10^5$--$10^{6}$ yr (for $n_{\rm ISM} = 0.1$--$1~\rm{cm}^{-3}$) prior to the explosion, probing the time-averaged ionizing luminosity of the progenitor over a significant part of its accretion history.
 
In this Letter, we report \textit{HST} non-detections, and upper limits, on the brightness of any extended H$\beta$ and \OIII\ emission nebula in pre- and post-explosion images at the site of SN~2014J (Section~\ref{sec:obs}). We then translate these limits in Section~\ref{sec:limmag} to new constraints on the progenitor system of SN~2014J. For an ambient ISM density of 1~$\rm{cm}^{-3}$, we exclude a hot luminous progenitor consistent with known SSSs in the past $\sim10^{5}$ yr prior to explosion, although for lower ISM densities we cannot exclude such a progenitor. In Section \ref{sec:discuss}, we discuss these limits in the context of previous efforts to constrain SN Ia progenitors and outline further near-term observations that can unambiguously rule out (or detect) any low density relic nebula associated with the progenitor of SN 2014J.
 

\section{Observations}
\label{sec:obs}

\begin{figure*}
\begin{center}
 \includegraphics[width=0.8\textwidth]{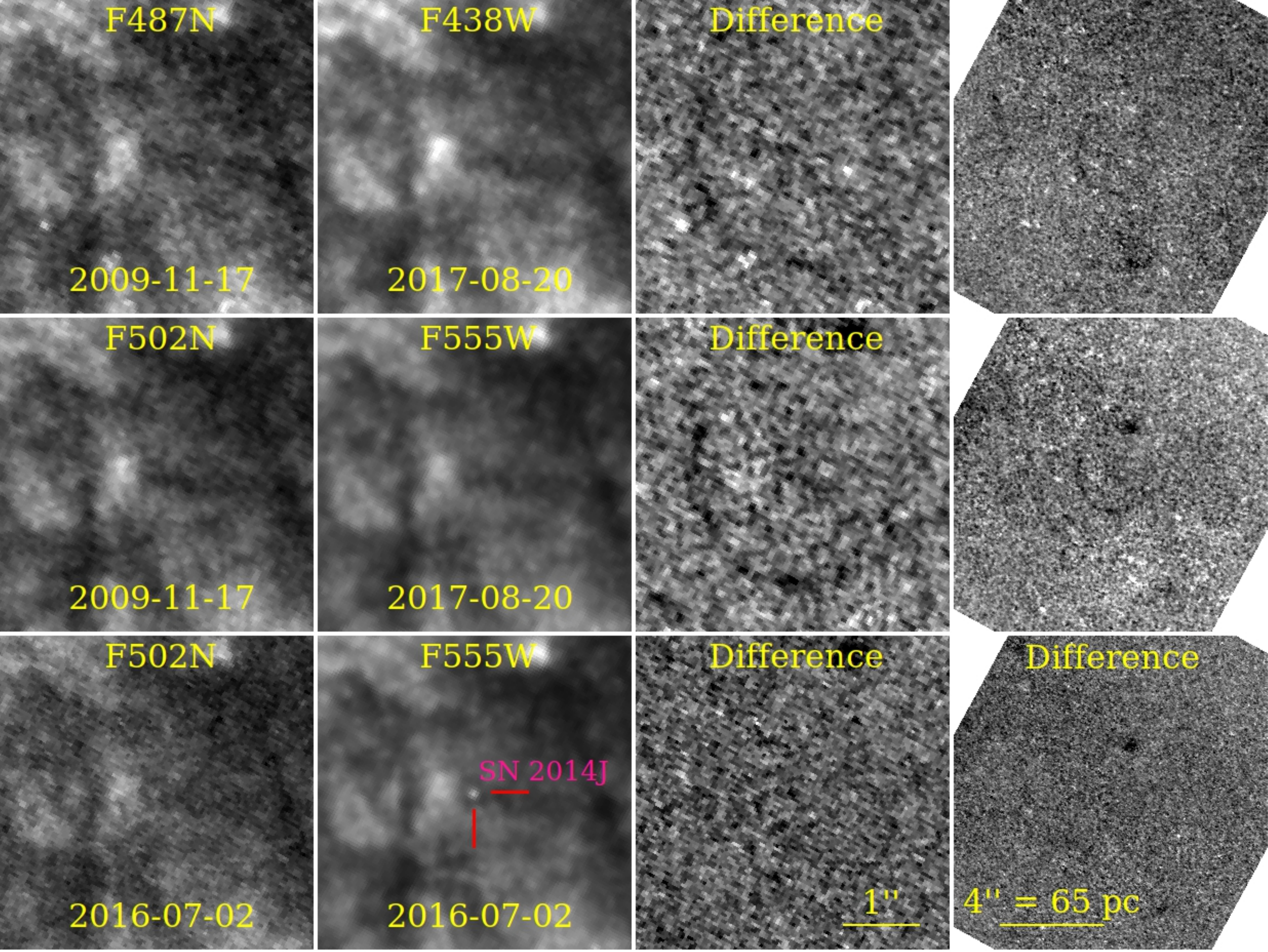}
 \end{center}
 \caption{The location of SN~2014J in F487N and F502N on 17 Nov. 2009 (top and centre row, respectively) and F502N on 02 Jul. 2016 (bottom row). From left to right, the columns show the narrow-band images, template wide-band images used to subtract the continuum, and the resultant difference images at two different resolutions, as marked. The SN is only visible in the F555W image taken on 02 Jul. 2016, where it is marked with a pink reticle. The SN light echo is apparent as a dark arc in the pre-explosion F487N and F502N difference images (top and centre right, respectively); the light echo and SN are both cleanly subtracted in the last F502N difference image. Each panel in the first three columns is roughly 4 arcsec (or 65 pc) on a side, while the fourth column is $\approx 6$ arcsec (or 200 pc) on a side; north is up and east is left.}
 \label{fig:loc}
\end{figure*}

Here, we use observations of SN 2014J with the Wide Field Camera 3 narrow-band filters F487N and F502N, as well as the broad-band filters F438W and F555W from \textit{HST} programs GO--11360 (PI: O'Connell), GO--14146 (PI: Crotts), and GO--14700 (PI: Sugerman). The location of the SN was imaged in both F487N and F502N on 17 Nov. 2009, $1536.9$ days before the SN reached maximum $B$-band light \citep{2015ApJ...798...39M}, with identical exposure times of 2455 s. The SN itself was also imaged in F502N on 02 Jul. 2016, $881.5$ days post maximum, with an exposure time of 2500 s.

In order to isolate the \Hb\ and \OIII\ emission lines from the continua captured by the filters, we scale and subtract broad-band F438W and F555W images taken at the same location, respectively. For the F502N image taken on 02 Jul. 2016, we use a concurrent F555W image with an exposure time of 1720 s. For the pre-explosion F487N and F502N images we use F438W (2600 s) and F555W (1664 s) images, respectively, taken on 20 Aug. 2017, when the SN is no longer detected. The resultant difference images are shown in Figure~\ref{fig:loc}. All images were first aligned using the IRAF tasks {\sc xregister} and {\sc wcscopy}.

SN 2014J is known to suffer from a large amount of extinction due to host-galaxy reddening. The line-of-sight Galactic extinction towards the SN in F487N and F502N is $0.505$ and $0.487$ mag, respectively, based on an interpolation of the values for nearby broad-band filters measured by \citet{2011ApJ...737..103S}. Adopting $A_V=1.8$ mag and $R_V$=1.46 for the host-galaxy extinction at the site of SN 2014J, as measured by \citet{2015ApJ...798...39M}, and assuming a \citet{1989ApJ...345..245C} reddening law, we compute host-galaxy extinctions of $2.4$ and $2.2$ mags, respectively.


\section{Analysis}
\label{sec:limmag}

\subsection{Constraints on any putative nebula}

We find no apparent extended source of \OIII\ or H$\beta$ emission at the location of SN 2014J, as shown in Figure~\ref{fig:loc}. To measure detection limits on the fluxes of these emission lines, we compute what surface brightness an extended source of a given radius would need to exhibit in order to achieve a signal-to-noise (S/N) ratio of 3 given the background noise level of each image. 

To measure the $3\sigma$ upper limit in each filter, we first use three methods to estimate the background noise: (1) using a fixed sky annulus $24.7$--$26$ pc away from the SN, just inside the reach of the surrounding light echo \citep{2015ApJ...804L..37C}; (2) using a variable sky annulus with a width of two pixels and spaced 4 pixels away from the aperture; and (3) using the target aperture itself. In all cases, the standard deviation and median of the pixel values within the specific aperture are treated as the background noise and median sky value. For the latter, we use the median instead of the mean to counteract the effect of point sources within the aperture. All three methods produce consistent results but the last two take into account that the background noise level slowly grows as the aperture begins to envelop regions of nebulosity at the outskirts of the image. Throughout this work we use the background noise as measured with the third method, since it produces the most conservative upper limits.

\begin{figure}
  \includegraphics[width=0.477\textwidth]{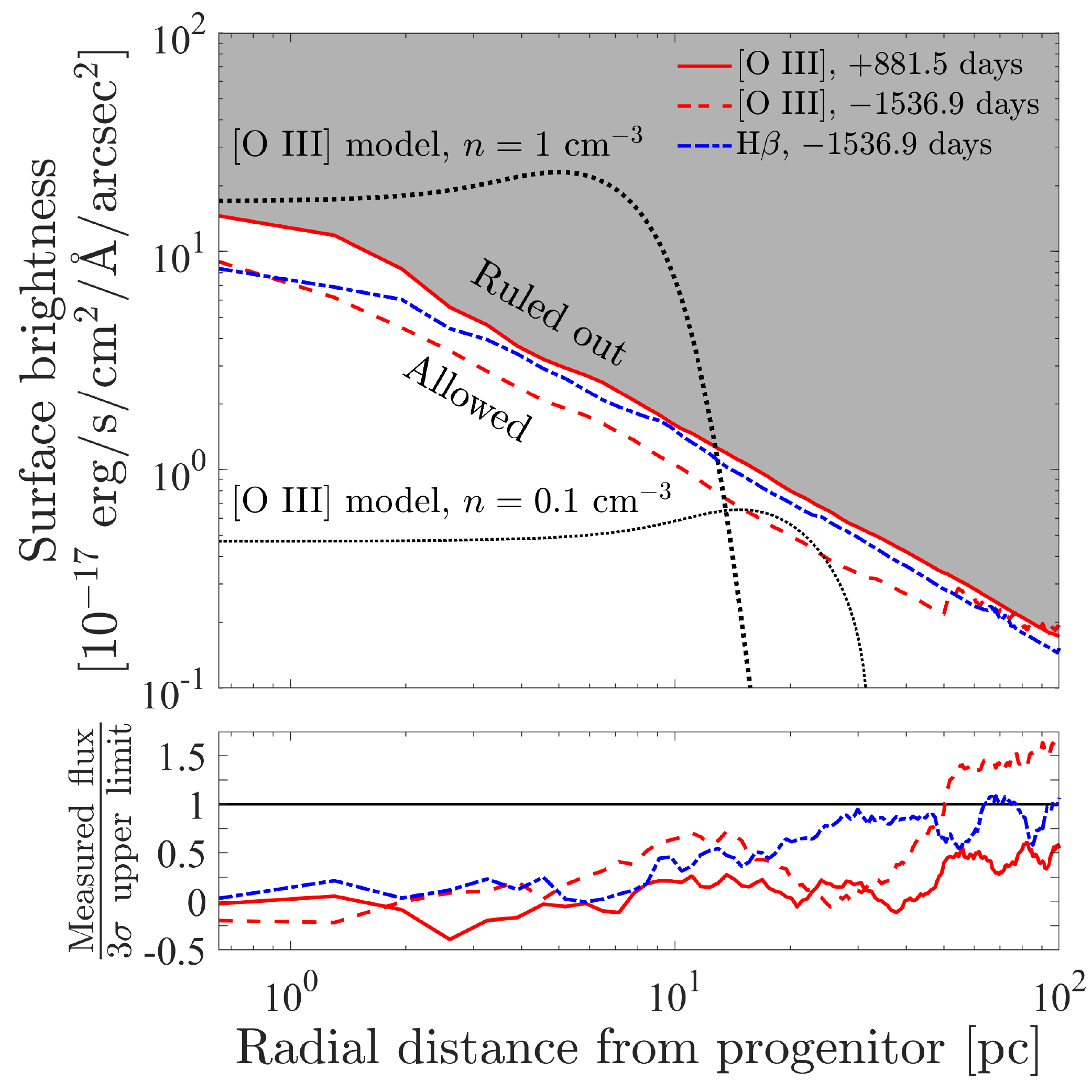}
 \caption{Top: Surface brightness of an emission nebula at the location of SN 2014J as a function of its radius. No such nebula is detected at the location of the SN. The $3\sigma$ upper limits on the surface brightness of such a nebula for \OIII\ at $881.5$ days past maximum light, and \OIII\ and H$\beta$ at $-1536.9$ days, are shown as the red solid, red dashed, and blue dashed-dotted curves, respectively. The gray shaded area shows the region of phase-space ruled out by the observations. This includes the predicted \OIII\ surface brightness of a source with ${\rm log}(L/{\rm erg~s^{-1}})=37.5$, $T=5\times10^5$~K, and $n_{\rm ISM}=1~{\rm cm}^{-3}$ (thick dotted curve). A similar source embedded in an ISM with $n_{\rm ISM}=0.1~{\rm cm}^{-3}$ (thin dotted curve) is allowed. Bottom: Ratio between the measured flux within an aperture and the estimated flux required for a signal with a S/N ratio of 3, as a function of the aperture radius. There is no significant signal in the pre-explosion H$\beta$ image or the post-explosion \OIII\ image. Beyond $\approx 50$ pc, the pre-explosion \OIII\ signal climbs above S/N$=3$, but we attribute this to diffuse sources to the SW and NW of SN 2014J, which begin to enter the aperture at that radius.}
 \label{fig:lims}
\end{figure}

With the background noise measured, the S/N ratio of the nebula is calculated as in the IRAF {\sc phot} routine. The flux is varied until the S/N reaches a value of 3. The resultant flux is then corrected for Galactic and host-galaxy extinction and divided by the area of the nebula, in square arcsec, to arrive at the final estimate of the $3\sigma$ upper limit on the nebula's surface brightness. The $3\sigma$ upper limits for all three difference images, as a function of the radius of the nebula, are shown in the upper panel of Figure~\ref{fig:lims}, where the gray shaded region denotes the area of phase-space ruled out by the observations.

The bottom panel of Figure~\ref{fig:lims} shows, as a function of distance from the location of SN 2014J, the ratio between the flux measured within the aperture and the $3\sigma$ upper limits computed above. No diffuse emission is detected at the location of SN 2014J (i.e., the ratio stays $<1$) out to $\approx 50$ pc. At this point, flux from diffuse sources to the NW and SW of the SN begin to enter the aperture. In the pre-explosion \OIII\ image, this causes the ratio to rise above 1, mimicking a S/N$\approx 4.5$ source. A similar rise is seen in the post-explosion \OIII\ image and, at $\approx 20$ pc, in the pre-explosion H$\beta$ image. In the latter, however, the contaminating flux is not enough to raise the total signal above a S/N ratio of 3. Taken together, we conclude that there is no \OIII\ or H$\beta$ diffuse emission centred on the location of SN 2014J out to at least 100 pc.

\subsection{Photoionization models using {\sc cloudy}}

In order to interpret our upper limits on the \OIII\ and H$\beta$ surface brightness in the context of a hot, luminous progenitor scenario, we must model the ionization state and nebular emission of the ISM given different plausible WD luminosities and temperatures. To do so, we employ the detailed photoionization and spectral synthesis code {\sc cloudy} (v13.03). For a given ISM density and metallicity, illuminating source, and geometry, {\sc cloudy} computes the ionization state and spectral emission of an arbitrary nebula by solving the equations of statistical and thermal equilibrium in 1-D. The development of {\sc cloudy} draws upon a substantial body of previous work; most relevant to the problem studied here, the code utilizes tables of recombination coefficients from \citet{Badnell03} and \citet{Badnell06}, with ionic emission data taken from the CHIANTI collaboration database version 7.0 \citep{Dere97,Landi12}.

In the following, we assume spherical symmetry and a surrounding ISM of uniform density and composition. Consistent with previous observations, we assume Solar metallicity \citep[e.g.,][]{McLeod93} and two fiducial models for the ISM density: $n_{\rm ISM} = 1~\rm{cm}^{-3}$ and $0.1~\rm{cm}^{-3}$, roughly bracketing the inferred ISM densities for most nearby, resolved SN Ia remnants (see discussion in Section~\ref{sec:intro}). Given this, we use {\sc cloudy} to compute the volumetric emissivities for both \OIII\ and H$\beta$ emission as a function of radius for varying plausible WD luminosities and temperatures, and then integrate along the line of sight to find the radial surface brightness profiles, as discussed by \cite{2016MNRAS.455.1770W}. As an illustrative example, two such \OIII\ radial profiles are shown in Figure~\ref{fig:lims} for each fiducial ISM density and an accreting, nuclear-burning WD luminosity and temperature consistent with the single-degenerate channel (${\rm log}(L/{\rm erg~s^{-1}})=37.5$ and ${\rm log}(T/{\rm K})=5.5$).  
Our results exclude the $n_{\rm ISM}=1~{\rm cm}^{-3}$ model, but do not preclude the model with $n_{\rm ISM}=0.1~{\rm cm}^{-3}$, as shown in Figure~\ref{fig:lims}.   

\begin{figure}
 \includegraphics[width=0.477\textwidth]{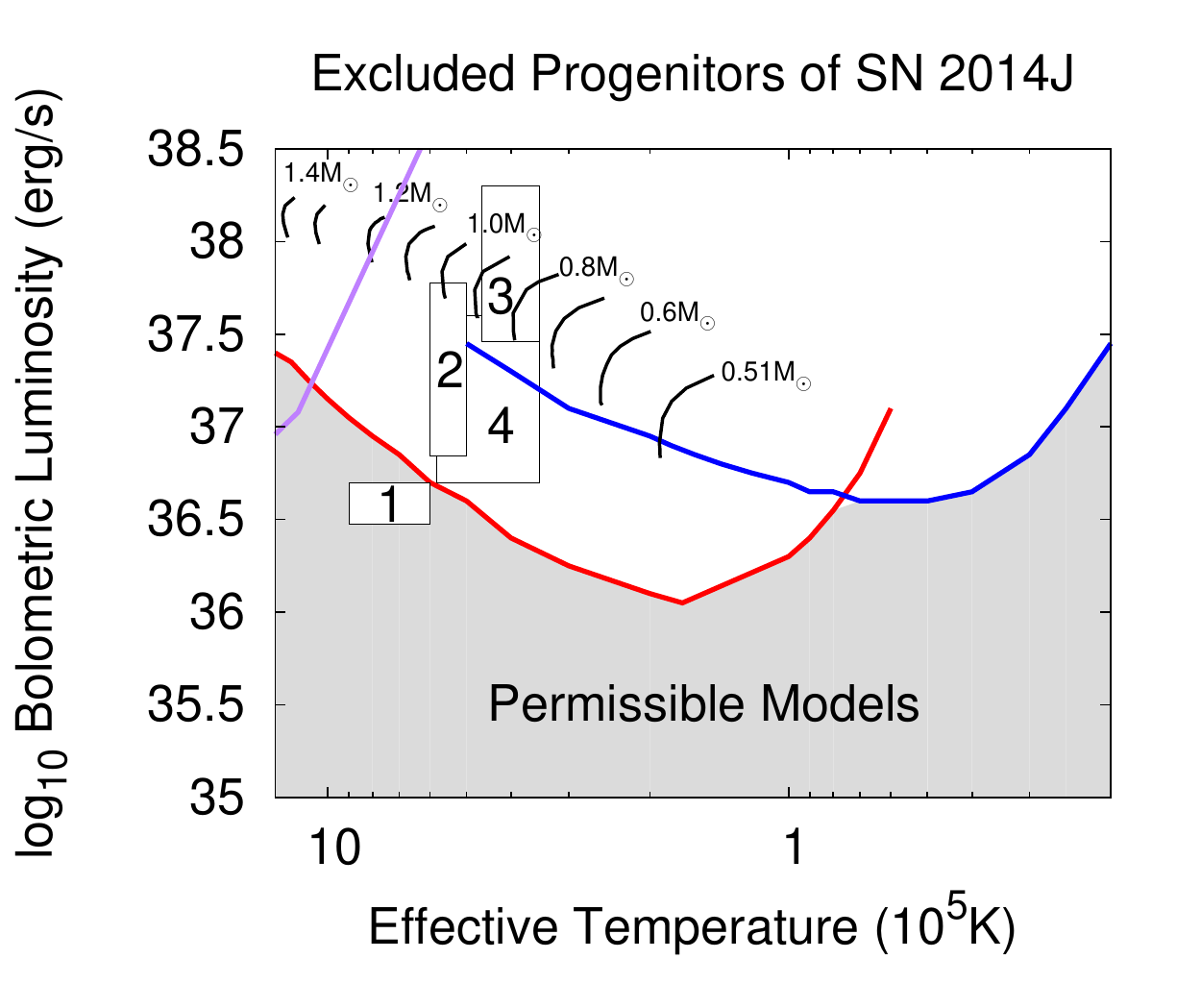}
\caption{Upper limits on the temperature and luminosity of the progenitor of SN 2014J, for a surrounding ISM density $n_{\rm ISM}=1~\rm{cm}^{-3}$ from the absence of detected \OIII\ $\lambda5007$~\AA\ (red line) and H$\beta$ emission (blue line). The shaded region denotes the permissible region of parameter space. The dashed purple line denotes previous X-ray upper limits from {\it Chandra} pre-explosion images. Thick black lines denote theoretical accreting stable nuclear-burning WD models, taken from \protect\cite{2013ApJ...777..136W}. Thin black boxes mark the approximate temperatures and luminosities inferred for confirmed close-binary Magellanic SSS with definitive luminosity measurements: 1.~CAL 87; 2.~1E 0035.4-7230; 3.~RX J0513.9-6951; and 4.~CAL 83. Note however that CAL 87 is viewed edge-on (i.e., obscured by its disk), and thought to be much more intrinsically luminous (see text).}
 \label{fig:constraint}
\end{figure}

In order to generalize this statement, we can then search for the minimum luminosity, at a given effective temperature, for which a putative SN Ia progenitor would produce a photoionization nebula with a sufficiently high \OIII\ surface brightness to violate our constraint. The result of this exercise is plotted in Figure~\ref{fig:constraint} for the $n_{\rm ISM}=1~\rm{cm}^{-3}$ ambient ISM density case. Also shown for reference are theoretical models for accreting, nuclear-burning WDs (within the so-called ``stable-burning strip'') from \cite{2013ApJ...777..136W}, as well as observed close-binary supersoft X-ray sources with well-determined luminosities and temperatures \citep{2000NewA....5..137G,2004ApJ...612L..53S}. 

From this comparison, it is clear that any plausible steadily nuclear-burning WD progenitor model is excluded for SN2014J; for most of the relevant phase space, the discrepancy is greater than 1--2 orders of magnitude. A notable exception appears to be CAL 87 (object ``1'' in Figure~\ref{fig:constraint}), which also appears to be inconsistent with theoretical models. This source, however, is viewed very nearly edge-on, and the disk is understood to at least partly obscure the central hot object \citep{2013A&A...559A..50N}. Therefore, it is almost certainly much more intrinsically luminous. We therefore conclude that, depending on the density of the surrounding ambient ISM, the progenitor of SN 2014J was not a SSS for a significant fraction of the $10^{5}$--$10^{6}$ yr prior to explosion.

Independent of the nuclear-burning luminosity, the accretion disk surrounding a sufficiently compact and rapidly-accreting WD may also be a significant source of UV and soft X-ray emission. Following \cite{2017NatAs...1..800W}, we model the disk spectrum assuming an optically thick, geometrically-thin Shakura-Sunyaev disk \citep{SS73}, with the inner boundary given by the WD radius, found from fitting the WD radius as a function of mass as determined by \cite{Panei00}. For varying WD mass and accretion rate, we may then use the disk spectra in our {\sc cloudy} simulations and carry out the same procedure as in the steady nuclear-burning case above. We find that for an approximately Chandrasekhar mass WD (here, we adopt the radius of a $1.35~{\rm M_{\odot}}$ WD), we may exclude accretion rates greater than $\sim 3\times10^{-7}~{\rm M_{\odot}}~{\rm yr}^{-1}$, based on the predicted \OIII\ surface brightness. This is above the limit for steady burning of hydrogen, where nuclear burning dominates the luminosity, and our previous result already excludes this regime. Our upper limit on the disk accretion rate, however, is independent of the composition of the matter being accreted. Therefore, this is constraining for WD progenitors with helium-donors; in particular, we may exclude Chandrasekhar-mass helium-accretors in the steady-burning or mild-flashes regimes \citep{Piersanti2014}.


\section{Discussion}
\label{sec:discuss}

The absence of a relic bright photoionization nebula surrounding SN 2014J strongly limits the ionizing luminosity of its progenitor during the last $\sim$100,000 years. In particular, for an assumed ISM density of $n_{\rm ISM}=1~\rm{cm}^{-3}$, we may exclude an unobscured, steadily nuclear-burning WD as the progenitor of SN 2014J. Dense winds from either the progenitor \citep[e.g., optically-thick winds, as in][]{1996ApJ...470L..97H} or its companion (e.g., a red giant), which could obscure an ionizing emission source, have already been excluded by radio and X-ray observations \citep{2014ApJ...790...52M,2014ApJ...792...38P}. This is consistent with other recent results excluding hot and highly luminous progenitors for individual nearby SNe and their remnants, as well as other more direct constraints on the high-energy progenitor luminosity from pre-explosion images \citep[e.g.,][]{2012MNRAS.426.2668N,2014MNRAS.442.3400N,2014MNRAS.442L..28G,2017NatAs...1..800W,2018ApJ...863..120W,2018MNRAS.481.4123K}. Similar to these other constraints, our study cannot exclude a long delay (here significantly greater than the recombination time) between a hot luminous phase and the explosion of SN 2014J, for instance due to the long spin-down time of a rapidly-rotating progenitor \citep[e.g.,][]{DiStefano2011,2011ApJ...730L..34J,2019MNRAS.482.5651M}. Any spin-up/spin-down scenario, however, faces a number of other challenges, as previously summarized in, e.g., \cite{2014ARA&A..52..107M}.

Notably, our upper limits may be directly compared with those found from pre-explosion archival {\it Chandra} images \citep{2014MNRAS.442.3400N}. In Figure~\ref{fig:constraint}, we show that for very high temperature sources, our upper limits are similar to those from the non-detection of soft X-rays, while for temperatures consistent with known supersoft sources, our limits are orders of magnitude deeper. Although our present results are not strongly constraining for lower ISM densities, any putative relic nebula associated with the progenitor would be accordingly more extended (with the Str{\" o}mgren radius scaling as $n^{-2/3}$). Given that such a nebula would remain for $10^{5}$--$10^{6}$ yr post-explosion, an ionizing progenitor embedded in low-density ISM can quickly be confirmed or excluded with deeper follow-up narrow-band imaging or integral-field spectroscopy (e.g., \citealt{2018arXiv181208799K}). This emphasizes the great utility of photoionization constraints on SN progenitors \citep{2016MNRAS.455.1770W}: rather than relying on serendipitous pre-explosion images, the influence of any luminous progenitor channel may be revealed even long after the explosion has faded.


\section*{Acknowledgments}
We thank Marat Gilfanov for helpful comments. O.G. is supported by an NSF Astronomy and Astrophysics Fellowship under award AST-1602595. This work is based on data obtained with the NASA/ESA \textit{Hubble Space Telescope}, all of which was obtained from MAST. This research has made use of the NASA/IPAC Extragalactic Database (NED) which is operated by the Jet Propulsion Laboratory, California Institute of Technology, under contract with NASA.


\end{document}